\documentclass[12pt]{article}

\usepackage{amssymb,amsmath,amsfonts,amsbsy}

\begin{document}

\title{Modular thermal inflation \\ without slow-roll approximation}

\author{
Jinn-Ouk Gong\footnote{jgong@kasi.re.kr}
\\ \\
{\em Department of Physics, KAIST, Daejeon, Republic of Korea}
\\ \vspace{0.5cm}
{\em International Center for Astrophysics, KASI, Daejeon, Republic of
Korea\footnote{Present address}}
}

\date{\today}

\maketitle

\begin{abstract}

We study an inflationary scenario where thermal inflation is followed by fast-roll
inflation. This is a rather generic possibility based on the effective potentials of
spontaneous symmetry breaking in the context of particle physics models. We show
that a large enough expansion could be achieved to solve cosmological problems.
However, the power spectrum of primordial density perturbations from the quantum
fluctuations in the inflaton field is not scale invariant and thus inconsistent with
observations. Using the curvaton mechanism instead, we can obtain a nearly scale
invariant spectrum, provided that the inflationary energy scale is sufficiently low
to have long enough fast-roll inflation to dilute the perturbations produced by the
inflaton fluctuations.

\end{abstract}

\vspace*{-80ex}

\hspace*{\fill} KAIST-TH/2006-03

\thispagestyle{empty}
\setcounter{page}{0}
\newpage
\setcounter{page}{1}

\section{Introduction}
\label{introduction}

Currently, inflation \cite{inf} is considered to be the most promising candidate to
provide the initial conditions for the successful hot big bang theory, solving many
cosmological problems such as homogeneity, isotropy and flatness of the observable
universe. At the same time, primordial density perturbations are generated from
quantum fluctuations, and they become the seeds of structure in the universe after
inflation. The most pristine form of these perturbations is inscribed as the
temperature anisotropy in the cosmic microwave background (CMB), which was first
probed by the Cosmic Background Explorer (COBE) satellite \cite{cobe}. Recently,
more improved CMB observations such as Wilkinson Microwave Background Probe (WMAP)
\cite{wmap} and BOOMERanG \cite{boomerang} detected the signature of the acoustic
oscillations in the anisotropy spectrum with unexperienced accuracy. Combined with
galaxy survey like Sloan Digital Sky Survey (SDSS) \cite{sdss}, these data strongly
support inflation.

There is, however, no consensus on the most plausible model of inflation. Many
conceptual developments in the inflationary scenario such as the idea of eternally
inflating universe \cite{eternal}, which suggests that inflation is a generic
feature in the early universe dominated by scalar fields and only the inflation of
the last 60 $e$-folds is relevant for the observed universe, have provided different
realisations of inflation, making our decision on the final stage of inflation even
more diverse. In that sense, the paradigm of slow-roll inflation \cite{slowroll} is
a very useful and attractive principle to discriminate which model is able to
implement long enough inflation for homogeneous and flat universe and to generate an
almost scale invariant spectrum of density perturbations. This helps us to clarify
which inflation model is viable by requiring that the inflaton potential $V(\phi)$
be flat enough to achieve the slow-roll conditions,
\begin{align}
\epsilon & = \frac{m_\mathrm{Pl}^2}{2} \left( \frac{V'}{V} \right)^2 \ll 1 \, ,
\nonumber \\
|\eta| & = \left| m_\mathrm{Pl}^2 \frac{V''}{V} \right| \ll 1 \, ,
\end{align}
where a prime denotes a derivative with respect to the inflaton field $\phi$ and
$m_\mathrm{Pl} = (8\pi G)^{-1/2} \simeq 2.4 \times 10^{18} \mathrm{GeV}$ is the
reduced Planck mass. But it is not easy to satisfy the slow-roll conditions, $|\eta|
\ll 1$ in particular, in many models motivated by particle physics. For example, in
supergravity theories the effective masses of generic scalar fields receive
corrections of $\mathcal{O}(H)$ during inflation, spoiling the condition $|\eta| \ll
1$. This does not mean, however, that inflation is impossible at all, and we could
obtain some inflation even when $\phi$ rolls off its effective potential quickly
\cite{fastroll}.

Also we can expect that the energy scale associated with the last inflationary stage
is considerably low compared with the Planck scale. This is also motivated by the
inflation models based on the de Sitter vacua construction by string moduli
stabilisation \cite{kklt}, where the Hubble parameter $H$ cannot be greater than
gravitino mass $m_{3/2}$ \cite{kkltH} which is of $\mathcal{O}(\mathrm{TeV})$ in
phenomenologically interesting gravity mediated supersymmetry breaking case. Such a
low scale inflation is also desirable to provide a solution to the cosmological
moduli problem \cite{moduli}. However, the well-known inflationary energy scale from
the observed magnitude of the density perturbations on the CMB scale is known as
\begin{equation}
V^{1/4} \simeq 2.77 \times 10^{-2} \epsilon^{1/4} m_\mathrm{Pl} \, .
\end{equation}
Also, we can obtain a similar bound from the contribution of the primordial
gravitational waves to the CMB anisotropy as \cite{gwbound}
\begin{equation}
V^{1/4} \simeq 3.0 \times 10^{-3} r^{1/4} m_\mathrm{Pl} \, ,
\end{equation}
where $r$ is the tensor-to-scalar amplitude ratio. Such a rather large inflationary
scale could be lowered by imposing some symmetry under which the inflaton $\phi$
transforms, so that inflation takes place near a symmetric point. Especially,
incorporating spontaneous breaking of the underlying symmetry, typically the
potential takes the form
\begin{equation}
V(\phi) \sim \lambda (\phi^2 - v^2)^2 \, ,
\end{equation}
where $v$ denotes the vacuum expectation value of $\phi$ and at the point $\phi =
0$, the top of the local maximum, the symmetry is preserved. It is then necessary
that $\phi$ is initially placed near the top of the effective potential, $\phi = 0$.
There are several ways to achieve it, and especially when the energy scale of
inflation is low, this could be implemented through thermal effects \cite{lowscale}.
This brings the idea of thermal inflation \cite{thermalinf} which takes place due to
the temperature corrections to the effective potential.

Thus, it is reasonable enough to consider inflation occurring near a maximum of the
effective potential, including thermal corrections, with significant curvature as
the inflation relevant for our observable universe, i.e., responsible for the
inflation of the last 60 $e$-folds. In this paper we are going to consider this
possibility; although this idea was suggested in Refs.~\cite{fastroll,dl2004}, our
discussion will be more explicit and detailed. This paper is outlined as follows. In
Section~\ref{inflation}, we first present the effective potential of our interest
and discuss the consequent inflationary phase. It consists of thermal and fast-roll
inflations, and we briefly describe their principles. In
Section~\ref{perturbations}, we discuss the density perturbations during inflation.
It is believed that the generation of perturbations is due to quantum fluctuations
of certain scalar field, which is usually expected to be the inflaton but could be
some different field, called the curvaton \cite{curvaton}. We will consider them
both. In Section~\ref{conclusions}, we summarise and conclude. Throughout this paper
we set $c = \hbar = 1$.

\section{Inflation}
\label{inflation}

An inflaton candidate of particular interest is a modulus field ubiquitous in string
theory \cite{modularinf}. Many moduli fields are expected to have Planckian vacuum
expectation values, with a potential of the form
\begin{equation}
V = M_\mathrm{SUSY}^4 \mathcal{F}(\phi/m_\mathrm{Pl}) \, ,
\end{equation}
where $M_\mathrm{SUSY}$ is the supersymmetry breaking scale, $\mathcal{F}$ is a
generic function whose typical values and derivatives are expected to be of
$\mathcal{O}(1)$, and $\phi$ is the scalar component of some relevant modulus field.
As discussed in the previous section, a class of potentials of particular interest
is the one associated with spontaneous symmetry breaking, and in this case inflation
may occur around a local maximum of the potential.

Hence we take the form of the potential, with a thermal correction term, as
\begin{equation}\label{thermalV}
V = V_0 + \left( g^2T^2 - \frac{1}{2}m_\phi^2 \right)\phi^2 + \cdots \, ,
\end{equation}
where $g$ is the coupling of $\phi$ to the fields of the background thermal bath,
and dots denote some unknown higher order function which gives the vacuum
expectation value of the inflaton at $\mathcal{O}(m_\mathrm{Pl})$, so that
\begin{equation}
V_0 \sim m_\phi^2m_\mathrm{Pl}^2 \, .
\end{equation}

\subsection{Thermal inflation}
\label{ti}

Thermal inflation \cite{thermalinf} was suggested as a solution to remove any
unwanted relics produced at the end of an earlier inflationary phase. Here we
briefly discuss the major principles of thermal inflation and refer the reader to
the original literatures \cite{thermalinf} for details.

When the potential is given as Eq.~(\ref{thermalV}), the universe is filled with
radiation and the inflaton. Then the energy density and the pressure are given by
\begin{align}
\rho & = \rho_T + V \, ,
\nonumber \\
p & = \frac{\rho_T}{3} - V \, ,
\end{align}
respectively, where $\rho_T = \pi^2g_*T^4/30$, with $g_*$ being the effective number
of relativistic degrees of freedom. Inflation takes place when $\rho + 3p < 0$, and
from above it reads $\rho_T < V$, i.e., when the potential dominates the total
energy density of the universe. Hence at the beginning of thermal inflation the
temperature is
\begin{equation}\label{Ti}
T_i \sim V_0^{1/4} \sim \sqrt{m_\phi m_\mathrm{Pl}} \, .
\end{equation}

Thermal inflation ends when the potential can no longer hold the inflaton at the
local minimum, and this happens when the effective mass squared becomes negative
so that instability develops at the origin; the effective mass squared is given
by
\begin{equation}\label{TI_meff}
m_\mathrm{eff}^2(T) = 2g^2T^2 - m_\phi^2 \, ,
\end{equation}
so the inflaton rolls away from the origin when temperature drops below
\begin{equation}\label{Tf}
T_f = \frac{m_\phi}{\sqrt{2}g} \, ,
\end{equation}
ending thermal inflation. Using $T \propto a^{-1}$, the number of the number of
$e$-folds during thermal inflation is estimated as
\begin{equation}
N_\mathrm{TI} \simeq \ln\left( \frac{T_i}{T_f} \right) \sim \ln\left(
\frac{V_0^{1/4}}{m_\phi} \right) \sim \frac{1}{2} \ln\left(
\frac{m_\mathrm{Pl}}{m_\phi} \right) \, .
\end{equation}
This alone is not enough to provide the observed homogeneous and isotropic universe
unless $m_\phi$ is vanishingly small, which does not seem very plausible in the
early universe. For example, taking $m_\phi \sim m_{3/2} \sim 10^3 \mathrm{GeV}$, it
gives $N_\mathrm{TI} \sim \ln 10^{15}/2 \sim 17$.

\subsection{Fast-roll inflation}
\label{fr}

After thermal inflation discussed in the previous section, the inflaton rolls
towards its minimum at $\mathcal{O}(m_\mathrm{Pl})$. At that moment, the slow-roll
parameter $|\eta| = |m_\mathrm{Pl}^2V''/V| \simeq m_\phi^2/(3H^2)$ is usually
constrained to be very small to maintain large number of $e$-folds and to obtain
nearly scale invariant spectrum. However, in many theories this condition is
violated, e.g., in supergravity theories, the soft masses of the scalar fields are
typically of $\mathcal{O}(H)$, making $|\eta| \sim 1$. Nevertheless, it is known
that still some amount of ``fast-roll" inflation could occur \cite{fastroll}.

We assume that throughout the fast-roll inflationary phase,
\begin{equation}\label{0tempV}
V = V_0 - \frac{1}{2}m_\phi^2\phi^2
\end{equation}
is a good enough approximation for the potential\footnote{Note that when this
assumption is not valid and higher order terms are responsible for ending inflation,
they may help us to build inflationary models with (very) low energy scale
\cite{lowscale}. Interestingly, still we can put almost the same observational
constraint for this general case \cite{hilltop}.}. Then the Hubble parameter is
practically constant, given by
\begin{equation}\label{frH}
H^2 = \frac{V_0}{3m_\mathrm{Pl}^2} \, .
\end{equation}
Now, solving the equation of motion
\begin{equation}
\ddot\phi + 3H\dot\phi + V' = 0 \, ,
\end{equation}
we obtain the solution as
\begin{align}
\phi(t) & = \phi_i \exp\left[ \left( \sqrt{\frac{9}{4} + \frac{m_\phi^2}{H^2}} -
\frac{3}{2} \right) Ht \right]
\nonumber \\
& = \phi_i e^{FHt} \, ,
\end{align}
where $\phi_i \sim \mathcal{O}(H)$ by the requirement that the classical motion of
$\phi$ be greater than the quantum fluctuation $H/(2\pi)$. This fast-roll inflation
ends\footnote{We can also postulate that inflation ends when the curvature becomes
$\mathcal{O}(1)$, which happens when $|\eta|_{\phi_f} = 1$. Then, $\phi_f$ is given
by
\begin{equation}
\phi_f = \sqrt{2}m_\mathrm{Pl} \sqrt{\frac{3H^2}{m_\phi^2} - 1} \, , \nonumber
\end{equation}
where we can see that there exists no real solution for $m_\phi^2 > 3H^2$. This
simply means $|\eta| > 1$ at the top of the potential so the curvature is always
greater than 1. We have, anyway, still $\phi_f \sim \mathcal{O}(m_\mathrm{Pl})$ when
$m_\phi^2 \sim \mathcal{O}(H^2)$.} when $\epsilon|_{\phi_f} = 1$, and this gives
\begin{equation}
\phi_f = \frac{m_\mathrm{Pl}}{\sqrt{2}} \left( \sqrt{1 + \frac{12H^2}{m_\phi^2}} - 1
\right) \, ,
\end{equation}
where we take $\phi_f > 0$. Hence when $m_\phi \sim \mathcal{O}(H)$, we have $\phi_f
\sim \mathcal{O}(m_\mathrm{Pl})$, i.e., fast-roll inflation holds until $\phi$
reaches its vacuum expectation value at $\mathcal{O}(m_\mathrm{Pl})$.

During fast-roll inflation, the universe expands with almost constant $H$ given by
Eq.~(\ref{frH}). The number of $e$-folds is then
\begin{equation}\label{FRefold}
N_\mathrm{FR} \sim F^{-1}\ln \left( \frac{m_\mathrm{Pl}}{H} \right) \sim
2F^{-1}\ln \left( \frac{m_\mathrm{Pl}}{V_0^{1/4}} \right) \, ,
\end{equation}
where we take $m_\phi^2 \sim \mathcal{O}(H^2)$ for the second approximation. To
solve the cosmological problems, we need at least $N_\mathrm{FR} \gtrsim 60 -
N_\mathrm{TI}$ after thermal inflation. This constrains the inflationary energy
scale and the inflaton mass, e.g., the intermediate scale $V_0^{1/4} \sim 10^{11}
\mathrm{GeV}$ gives a bound on the mass squared to be $m_\phi^2 \lesssim
\mathcal{O}(H^2)$ to obtain the total expansion of 60 $e$-folds. Some representative
values are shown in Table~\ref{parameters}.

\begin{table}[h]
\begin{center}
\begin{tabular}{llllll}
\hline
    $g$ & $m_\phi$ (GeV) & $H^2$ & $F$ & $N_\mathrm{TI}$ & $N_\mathrm{FR}$
    \\
\hline\hline
    1 & $10^{-3}$ & $3m_\phi^2$ & 0.107 & 24.96 & 453.79
    \\
    1 & $1$ & $m_\phi^2$ & 0.303 & 21.51 & 139.78
    \\
    0.1 & $10^3$ & $m_\phi^2/3$ & 0.791 & 15.75 & 45.45
    \\
    0.1 & $10^9$ & $m_\phi^2/10$ & 2 & 8.84 & 11.36
    \\
    0.01 & $10^{12}$ & $m_\phi^2/100$ & 8.61 & 3.09 & 1.97
    \\
\hline
\end{tabular}
\caption{\label{parameters} A few interesting parameter sets. Note that when
$m_\phi$ becomes heavier, or the inflationary energy scale $V_0^{1/4}$ gets higher,
we obtain smaller number of $e$-folds.}
\end{center}
\end{table}

\section{Perturbations}
\label{perturbations}

It is well known that during inflation, primordial density perturbations are
generated from quantum fluctuations of one or more scalar fields. These
perturbations later become the seeds of the formation of structure in the universe.
The adiabatic component is associated with the primordial curvature perturbation,
whose power spectrum is given by \cite{wmapP}
\begin{equation}\label{wmap_P}
\mathcal{P}^{1/2} \simeq 5 \times 10^{-5} \, ,
\end{equation}
and the spectral index is \cite{sdss,wmapP}
\begin{equation}\label{wmap_n}
n = 0.97 \pm 0.03 \, ,
\end{equation}
making the power spectrum nearly scale invariant on large observable scales. It is
usually believed that the quantum fluctuations of the inflaton result in the
primordial curvature perturbations. An interesting alternative, called the curvaton
scenario \cite{curvaton}, suggests that some scalar field different from the
inflaton is responsible for the generation of perturbations. In this section, we
explore both possibilities. Note that since we are interested in low inflationary
energy scale, the amplitude of the power spectrum of the primordial gravitational
waves will be suppressed to an unobservable level and we will not consider it here;
see, e.g., Ref.~\cite{gw} for a discussion on the spectrum and the spectral index
for the primordial gravitational waves.

\subsection{Inflaton case}
\label{inflatonpert}

\subsubsection{Thermal inflation}
\label{tipert}

During thermal inflation, from Eqs.~(\ref{thermalV}) and (\ref{TI_meff}), the
effective potential could be written as
\begin{equation}
V = V_0 + \frac{1}{2}m_\mathrm{eff}^2\phi^2 \, ,
\end{equation}
where $m_\mathrm{eff}^2 \sim m_\phi m_\mathrm{Pl}$ at the early stage of thermal
inflation, which is far larger than $H^2 \sim m_\phi^2$. The inflaton $\phi$ is,
therefore, well anchored at the false vacuum. In this case, the quantum fluctuations
of $\phi$ do not become classical perturbations\footnote{More exactly, for a generic
scalar field $\chi$ with $m_\chi
> 3H/2$, the fluctuations of $\chi$ do not produce classical perturbations. This
could be seen from the mode equation \cite{modeeq,sg2001}
\begin{equation}
u_\mathbf{k}'' + \left[ k^2 - \frac{1}{\tau} \left( \nu^2 - \frac{1}{4} \right)
\right] u_\mathbf{k} = 0 \, , \nonumber
\end{equation}
where $u_\mathbf{k} = a \delta\chi_\mathbf{k}$, $\tau = \int dt/a$ is the conformal
time, and
\begin{equation}
\nu^2 = \frac{9}{4} - \frac{m_\chi^2}{H^2} \, . \nonumber
\end{equation}
Only when $m_\chi < 3H/2$, $\nu$ becomes real and the well-known Hankel function
solution is obtained.}. The resulting power spectrum of the inflaton fluctuations is
given by \cite{heavyP}
\begin{equation}
\mathcal{P}_{\delta\phi} = \left( \frac{H_\star}{2\pi} \right)^2 \left(
\frac{k}{aH_\star} \right)^3 \exp \left( -\frac{2m_\mathrm{eff}^2}{H_\star^2}
\right) \, ,
\end{equation}
where $\star$ denotes the epoch of horizon crossing $k = aH$. This spectrum is not
scale invariant but strongly blue with the spectral index being equal to 4, and the
amplitude exponentially suppressed. At later stages of thermal inflation, however,
$m_\mathrm{eff}$ gets lighter and finally becomes smaller than $3H/2$, capable of
producing classical perturbations. What is the corresponding spectrum of the
primordial curvature perturbation? Its exact form is, unfortunately, not known yet.
Nevertheless, we can anticipate it in several ways; perhaps the simplest expectation
is that it is related to $\mathcal{P}_{\delta\phi}$ in a similar manner to the case
of the usual slow-roll inflation, so that $\mathcal{P}$ should be also blue. This
could be expected from the simple observation that since the background is de
Sitter, the quantum fluctuations of $\phi$ decay as they go outside the horizon.
Hence, the amplitude of those which exit earlier, i.e., on larger scales, is smaller
than those which exit later. This makes the spectrum blue. We can derive the same
conclusion from the argument that as one approaches $T_f$ the fluctuations will grow
bigger, since at the time when $m_\mathrm{eff}^2 = 0$ the effective potential is
constant, i.e., $V = V_0$, then the fluctuations become very large\footnote{In fact,
if the effective potential remains constant, the universe expands in the pure de
Sitter background, and the spectrum will be infinite. Even if $\phi$ is assumed to
be able to move on this constant potential (the so-called ultra-slow-roll inflation)
it will stop at some point, say $\phi_\mathrm{dS}$, making the power spectrum
infinite there. If we introduce a cutoff at $\phi_c$ before $\phi$ reaches
$\phi_\mathrm{dS}$, we have a scale invariant power spectrum \cite{kinney}. Note
that when inflation is not suspended after $\phi_c$, generally a (large) peak around
the scale corresponding to $\phi_c$ is expected in the spectrum \cite{bsi}. This is
somewhat similar to the situation we are discussing now, as we will see in the
following section.}. The spectrum therefore is blue during thermal inflation stage.

Also there is another source of perturbations. Thermal fluctuations during thermal
inflation may cause the fluctuations in the number of $e$-folds, leading to
curvature perturbation. That is, the perturbation in the curvature of the final
comoving hypersurfaces $\mathcal{R}$ is expressed as \cite{multifield}
\begin{equation}\label{thermalR}
\mathcal{R} = \delta N = \frac{\partial N}{\partial T}\delta T \, .
\end{equation}
When $T \gg H$, within a Hubble volume of radius $H^{-1}$, there exist
$H^{-3}/T^{-3}$ thermal baths of correlation length $T^{-1}$. Hence the typical
thermal fluctuation on the scale of $H^{-1}$ is
\begin{equation}
\delta T \sim \frac{T}{\sqrt{H^{-3}/T^{-3}}} \, ,
\end{equation}
and with Eq.~(\ref{thermalR}) this gives
\begin{equation}
\mathcal{R} \sim \left( \frac{H}{T} \right)^{3/2} \sim \left( \frac{T_f}{T}
\right)^{3/2} \, ,
\end{equation}
where we have used Eq.~(\ref{Tf}) with $g$ being of $\mathcal{O}(1)$. The
corresponding spectrum has a spectral index $n = 4$, i.e., steeply blue, with its
maximum amplitude of $\mathcal{O}(1)$ at the end of thermal inflation.

\subsubsection{Fast-roll inflation}
\label{frpert}

When the potential has the form of an inverted quadratic one as Eq.~(\ref{0tempV}),
the corresponding power spectrum is known as \cite{sg2001}
\begin{equation}\label{fr_P}
\mathcal{P} = \frac{V_0}{12\pi^2m_\mathrm{Pl}^2\eta^2\phi_i^2} \left[ 2^{-\eta}
\frac{\Gamma(3/2 - \eta)}{\Gamma(3/2)} \right]^2 \, ,
\end{equation}
and the spectral index as
\begin{equation}\label{fr_n}
n - 1 = 2\eta \, .
\end{equation}
Since the initial value of $\phi$ is of $\mathcal{O}(H)$ as estimated in
Section~\ref{fr}, when $m_\phi^2 \sim \mathcal{O}(H^2)$ so that $|\eta| \sim
\mathcal{O}(1)$, the amplitude of the spectrum is
\begin{equation}\label{frP_amp}
\mathcal{P} \sim \left( \frac{H}{\eta\phi_i} \right)^2 \sim \eta^{-2} \sim
\mathcal{O}(1) \, ,
\end{equation}
and the corresponding spectral index is
\begin{equation}
n \sim \mathcal{O}(-1) \, .
\end{equation}
Hence, at the beginning of fast-roll inflation the amplitude of the perturbation
spectrum is of $\mathcal{O}(1)$, and it decreases very quickly at later stages; that
is, we obtain a steep red spectrum.

Here a question may arise; is it possible to use thermal fluctuations to compensate
the red tilt and obtain a nearly flat spectrum? When thermal inflation ends and
fast-roll inflation begins, the temperature $T_f$ is of $\mathcal{O}(m_\phi) \sim
\mathcal{O}(H)$, as can be seen from Eq.~(\ref{Tf}). Once the stage of fast-roll
inflation sets in, the universe expands with almost constant $H$ given by
Eq.~(\ref{frH}), and accordingly temperature decreases exponentially. Hence an
entire Hubble volume is enclosed in a single thermal bath of correlation length
$T^{-1} \gg H^{-1}$, and the effects of thermal fluctuations are completely
negligible. Moreover, even if we could make $H \sim T$ for a long time, to
compensate the steep blue tilt $n = 4$ due to thermal fluctuations, we need a large
$m_\phi$. Such a large mass finishes fast-roll inflation very quickly, well before
total 60 $e$-folds. Unless some special mechanism or finely tuned condition is
assumed, it seems very difficult to make the spectrum flat.

The large perturbations produced at the early stage of fast-roll inflation may lead
to cosmological disasters. For example, if they are not swept away by the following
longer stage of inflation, they would cause an unacceptably copious black hole
production when inflation ends and the density of the universe is dominated by the
coherent scalar condensates, i.e., oscillating massive scalar fields which are
equivalent to non-relativistic matter.

\subsection{Curvaton case}
\label{curvatonpert}

In the curvaton scenario, during inflation, some scalar field other than the
inflaton, the curvaton field $\sigma$, is assumed to be almost free with small
effective mass, i.e., $|\partial^2\mathcal{V}/\partial\sigma^2| =
|\mathcal{V}_{\sigma\sigma}| \ll H^2$, where $\mathcal{V}$ is the curvaton
potential. The spectrum of the quantum fluctuations of $\sigma$ on superhorizon
scales is therefore given by
\begin{equation}
\mathcal{P}_{\delta\sigma} = \frac{H_\star}{2\pi} \, .
\end{equation}
The isocurvature perturbation associated with these fluctuations\footnote{In the
models of inflation involving several inflaton fields, we can obtain significant
isocurvature perturbations \cite{multiiso} as well as conventional curvature
perturbations \cite{multifield}.} later become curvature perturbation when the
curvaton oscillates at the minimum of its potential and decays. The corresponding
spectrum of the primordial curvature perturbation is \cite{curvatonpert}
\begin{equation}\label{curvaton_P}
\mathcal{P}^{1/2} = \frac{2}{3}rq \frac{H_\star}{2\pi\sigma_\star} \, ,
\end{equation}
where
\begin{equation}
r = \left. \frac{\rho_\sigma}{\rho} \right|_\mathrm{dec}
\end{equation}
is the ratio of the curvaton energy density to the total energy density of the
universe at the epoch of curvaton decay, and $q \lesssim 1$ is a constant. The
spectral index is given by
\begin{equation}\label{curvaton_n}
n - 1 = 2\eta_{\sigma\sigma} - 2\epsilon \, ,
\end{equation}
where $\eta_{\sigma\sigma} = \mathcal{V}_{\sigma\sigma}/(3H^2)$, the slow-roll
parameter with respect to $\sigma$, determines the value $n - 1$ in many physically
interesting classes of inflation models where $\epsilon$ is negligible. Therefore,
we can easily obtain a nearly scale independent, flat spectrum as long as
$|\eta_{\sigma\sigma}| \ll 1$.

\subsubsection{Thermalisation}
\label{thermalise}

One thing we should make sure at this stage is that the curvaton should not join the
surrounding thermal bath before its oscillation commences. Otherwise, there is no
chance for the curvaton to decay into other particle species and it becomes simply a
component of the thermal bath, and scales as radiation, i.e., $\rho_\sigma \propto
T^4$. Let the effective mass squared of the curvaton consist of the soft mass and
the thermal correction as Eq.~(\ref{TI_meff}),
\begin{equation}
\mathcal{V}'' \sim m_\sigma^2 + g'^2T^2 \, .
\end{equation}
Then, to avoid thermalisation, we demand that \cite{thermalize}
\begin{equation}
g'T_m < m_\sigma \, ,
\end{equation}
where $T_m$ is the temperature when the effective curvaton mass becomes dominated by
the soft mass $m_\sigma$. The highest temperature of our interest is $T_i$ at the
beginning of thermal inflation given by Eq.~(\ref{Ti}), and this gives
\begin{equation}\label{therm_bound}
m_\sigma > g' \sqrt{m_\phi m_\mathrm{Pl}} \, .
\end{equation}
From Eq.~(\ref{wmap_n}), we obtain $|m_\sigma| \lesssim 0.2 H$ provided that the
soft mass of the curvaton is completely dominating the effective mass. Then,
combining with Eq.~(\ref{therm_bound}), we can find an upper bound on the coupling
$g'$ as\footnote{Note that more rigorous bounds on the coupling of the curvaton to
the thermal bath are given in Ref.~\cite{thermalize} in various situations. It is,
however, interesting that we can derive a similar bound using a very simple
argument.}
\begin{equation}
g' \lesssim 0.2 \sqrt{\frac{H}{m_\mathrm{Pl}}} \, .
\end{equation}
If we take $H \sim 10^3 \mathrm{GeV}$, this gives $g' \lesssim 10^{-8}$; the
curvaton is therefore required to hardly interact with the thermal bath indeed.

\subsubsection{Suppressing the inflaton perturbations}
\label{suppressing}

From the discussion of Section~\ref{inflatonpert}, we found that the perturbation
spectrum associated with the inflaton is highly scale dependent. Especially, the
large peak of amplitude of $\mathcal{O}(1)$ at the transition between the thermal
and the fast-roll inflationary stages seems unavoidable. Meanwhile, in the curvaton
scenario it is assumed that initially the universe is unperturbed practically. This
means the curvature perturbation originated from the fluctuations in the inflaton is
negligible, leaving only isocurvature perturbation at the end of inflation. Hence,
it is necessary to suppress the curvature perturbation associated with the inflaton
for the curvaton scenario to work properly.

One obvious way of achieving this is to have a long enough period of fast-roll
inflation. Since the spectrum is steeply red, i.e., the amplitude of the
perturbation produced at later stages of fast-roll inflation is much smaller, soon
we approach the universe with negligible curvature perturbation, suitable for
implementing the curvaton scenario. From the number of $e$-folds during fast-roll
inflation, Eq.~(\ref{FRefold}), it is clear that the inflationary energy scale is
{\em demanded} to be low to have large $N_\mathrm{FR}$. By defining $x = -k\tau =
k/(aH)$, we obtain
\begin{equation}\label{dNdlnx}
dN = H dt = -d \ln x \, .
\end{equation}
From Eqs.~(\ref{fr_P}) and (\ref{fr_n}), we can write the spectrum simply as
\begin{equation}
\mathcal{P}^{1/2} = \sqrt{\frac{V_0}{12\pi^2m_\mathrm{Pl}^2\eta^2\phi_i^2}}
2^{-\eta} \frac{\Gamma(3/2 - \eta)}{\Gamma(3/2)} x^\eta = A x^\eta \, .
\end{equation}
Here, we can set $x = 1$ at the beginning of fast-roll inflation so that
$\mathcal{P}^{1/2}|_{x = 1} = A \sim \mathcal{O}(1)$ when $|\eta| \sim
\mathcal{O}(1)$, as was already noted from Eq.~(\ref{frP_amp}). Then, let $x =
x_\mathrm{curv}$ when the amplitude of the spectrum becomes small enough for the
curvaton scenario to work properly, say, of $\mathcal{O}(10^{-10})$. That is,
\begin{equation}
\mathcal{P}^{1/2}|_{x = x_\mathrm{curv}} = A x_\mathrm{curv}^\eta \sim
\mathcal{O}(10^{-10}) \, ,
\end{equation}
so $x_\mathrm{curv} = (10^{-10}/A)^{1/\eta}$. Then, from Eq.~(\ref{dNdlnx}) the
number of $e$-folds between $x_0$ and $x_\mathrm{curv}$ is simply
\begin{equation}
\Delta N = -\ln x_\mathrm{curv} = -\eta^{-1} \ln \left( \frac{10^{-10}}{A} \right)
\, .
\end{equation}
Therefore the required number of $e$-folds during fast-roll inflation should be
greater than $60 + \Delta N$; 60 $e$-folds necessary to solve various cosmological
problems, and $\Delta N$ to dilute the perturbation associated with the inflaton so
that the curvaton scenario can work. Combining this with Eq.~(\ref{FRefold}), we
find
\begin{equation}
2F^{-1} \ln \left( \frac{m_\mathrm{Pl}}{V_0^{1/4}} \right) \gtrsim 60 - \eta^{-1}
\ln \left( \frac{10^{-10}}{A} \right) \, .
\end{equation}
By estimating $x_\mathrm{curv} \sim 10^{-10}$, we obtain $\Delta N \simeq 23$. This
gives
\begin{equation}
V_0^{1/4} \lesssim \exp \left( -\frac{83}{2}F \right) m_\mathrm{Pl} \, .
\end{equation}
The heavier $m_\phi$ gets, the tighter this bound becomes. For example, when
$m_\phi^2 = H^2$, this gives a rather mild bound of $V_0^{1/4} \lesssim 8.38 \times
10^{12} \mathrm{GeV}$. Instead, if $m_\phi^2 = 3H^2$, we find $V_0^{1/4} \lesssim
1.31 \times 10^5 \mathrm{GeV}$.

\subsubsection{Curvaton dominance in low inflationary scale}
\label{lowscale}

In the previous subsection, we have seen that to implement the curvaton scenario
successfully, we need a sufficiently low inflationary energy scale to dilute the
perturbations associated with the inflaton fluctuations. With such a low scale,
however, it is not possible to generate the observed magnitude of density
perturbations, given by Eq.~(\ref{wmap_P}). Indeed, in the simplest curvaton model,
the inflationary Hubble parameter $H_\star$ is required to be greater than $10^7
\mathrm{GeV}$ \cite{curvatonlowscale}. Using the bound
\begin{equation}
r \lesssim \frac{\sqrt{m_\sigma m_\mathrm{Pl}}
\sigma_\star^2}{T_\mathrm{dec}m_\mathrm{Pl}^2} \, ,
\end{equation}
the constraint $m_\sigma < H_\star$ and the big bang nucleosynthesis bound
$T_\mathrm{dec} > 1 \mathrm{MeV}$, $H_\star > 10^7 \mathrm{GeV}$ is translated into
\begin{equation}\label{ini_amp}
\sigma_\star \gtrsim 5.54 \times 10^{10} \mathrm{GeV} \, .
\end{equation}
Combining Eqs.~(\ref{curvaton_P}) and (\ref{wmap_P}), we obtain a relation
\cite{curvatonlowscale}
\begin{equation}\label{ini_amp2}
\sigma_\star \simeq 2 \times 10^3 r H_\star.
\end{equation}
When the inflationary energy scale is low, e.g., $H_\star \sim
\mathcal{O}(\mathrm{TeV})$, we cannot satisfy Eq.~(\ref{ini_amp}) hence the
amplitude of the power spectrum is inconsistent with observations.

This is equivalent to the absence of the curvaton dominated universe; for the
curvaton to dominate the energy density of the universe before its decay, we
need\footnote{Note that in the curvaton scenario we are considering, the curvaton
$\sigma$ begins oscillation after the universe is filled with radiation due to the
decay of the inflaton. The decay of the curvaton happens after its oscillation
commences. Hence, we have
\begin{equation}
\Gamma_\sigma < m_\sigma < \Gamma_\phi \, , \nonumber
\end{equation}
where $\Gamma_\sigma$ and $\Gamma_\phi$ denote the decay rate of the curvaton and
the inflaton, respectively. When this condition is satisfied, $\rho_\sigma$
decreases as $a^{-3}$ in the background thermal bath.} \cite{mt2002}
\begin{equation}
\left( \frac{\Gamma_\sigma}{m_\sigma} \right)^{1/4} \lesssim 2 \times 10^3 r
\frac{H_\star}{m_\mathrm{Pl}} \, ,
\end{equation}
where we have used Eq.~(\ref{ini_amp2}). If the curvaton were to dominate the energy
density ($r = 1$) given a low inflationary scale ($H_\star \sim 10^3 \mathrm{GeV}$),
the decay rate of the curvaton is estimated to be $\Gamma_\sigma \lesssim 10^{-48}
m_\sigma$. Then, using $m_\sigma < H_\star$, the reheating temperature after the
decay of the curvaton is
\begin{equation}
T_\mathrm{R'} \sim g_*^{-1/4} \sqrt{\Gamma_\sigma m_\mathrm{Pl}} \lesssim 10^{-15}
\mathrm{GeV} \, ,
\end{equation}
which is far below the big bang nucleosynthesis bound 1 MeV. Hence, at low
inflationary energy scale, the curvaton cannot dominate the universe and is unable
to produce the observed magnitude of the perturbation spectrum. To overcome this
difficulty, several alternatives were suggested, e.g., the case of the curvaton as a
pseudo Nambu-Goldstone boson with a varying order parameter to amplify the curvaton
perturbations \cite{pngbcurvaton}.

\section{Conclusions}
\label{conclusions}

Despite many attractive features of the inflationary universe, it is not trivial to
construct the inflation model responsible for the observable universe, i.e., the
inflation of the last 60 $e$-folds. One obvious difficulty is that it is not easy to
achieve the slow-roll conditions in the context of particle physics models, e.g., in
supergravity theories. This makes the total expansion of the universe during the
stage of inflation very short and the spectrum of the primordial curvature
perturbation produced during inflation highly scale dependent. Also the inflationary
energy scale expected from observations is rather large, causing the troublesome
moduli problem after inflation.

In this paper, we have considered a simple model free from such constraints. By
imposing some symmetry under which the inflaton transforms, we can obtain an
effective potential which describes a local maximum and lower the inflationary
energy scale considerably. The inflaton could be placed near such a local maximum
via thermal effects when the energy scale is low. Then, the consequent inflation
consists of two phases, thermal inflation and fast-roll inflation. The total number
of $e$-folds could be large enough to solve cosmological problems provided that the
energy scale of inflation is sufficiently low. The power spectrum of the primordial
curvature perturbation from the quantum fluctuations in the inflaton is, however,
highly scale dependant, inconsistent with observations. This could be evaded by
adopting the curvaton scenario where the curvature perturbation is produced from the
isocurvature perturbation of a light scalar field different from the inflaton, the
curvaton. For the curvaton scenario to work properly, the inflationary energy scale
is required to be low enough to maintain the fast-roll inflation long enough to
dilute the perturbation from the inflaton fluctuations away.

\subsection*{Acknowledgements}
I am deeply indebted to William Kinney, David Lyth, Ewan Stewart and especially
Misao Sasaki for invaluable comments, discussions and suggestions. Also I thank the
anonymous referee for important remarks on earlier drafts. This work was supported
in part by the Brain Korea 21.

\end{document}